\documentclass{article}

\usepackage{arxiv}

\usepackage[utf8]{inputenc} 
\usepackage[T1]{fontenc}    
\usepackage{hyperref}       
\usepackage{url}            
\usepackage{booktabs}       
\usepackage{amsfonts}       
\usepackage{nicefrac}       
\usepackage{microtype}      
\usepackage{graphicx}
\graphicspath{ {./images/} }

\usepackage{xcolor}
\usepackage{amsfonts, amsmath, amssymb}
\usepackage{listings}
\usepackage{xcolor}
\usepackage{ulem}

\DeclareEmphSequence{\itshape}

\def\EE{{\cal E}} \def\VV{{\cal V}}

\def\GG{{\cal G}}

\definecolor{codegreen}{rgb}{0,0.6,0}
\definecolor{codegray}{rgb}{0.5,0.5,0.5}
\definecolor{codepurple}{rgb}{0.58,0,0.82}
\definecolor{backcolour}{rgb}{0.95,0.95,0.92}

\title{\texttt{adabmDCA 2.0} – a flexible but easy-to-use package \\ for Direct Coupling Analysis}

\author{
 Lorenzo Rosset \\
  Laboratory of Computational and Quantitative Biology \\
  Sorbonne Universit\'e, CNRS\\
  75005 Paris, France\\
  and \\
  Laboratoire de Physique Théorique \\
  École Normale Supérieure \\
  75231 Paris, France \\
  \texttt{lorenzo.rosset@phys.ens.fr} \\
   \And
 Roberto Netti \\
  Laboratory of Computational and Quantitative Biology \\
  Sorbonne Universit\'e, CNRS\\
  75005 Paris, France\\
  \And
 Anna Paola Muntoni \\
  DISAT \\
  Politecnico di Torino \\
  10129 Torino, Italy
  \And
 Martin Weigt \\
  Laboratory of Computational and Quantitative Biology \\
  Sorbonne Universit\'e, CNRS\\
  75005 Paris, France\\
  \And
 Francesco Zamponi \\
  Dipartimento di Fisica \\
  Sapienza Universit\`a di Roma \\
  00185 Rome, Italy
}

\begin{document}
\maketitle
\begin{abstract}
In this methods article, we provide a flexible but easy-to-use implementation of Direct Coupling Analysis (DCA) based on Boltzmann machine learning, together with a tutorial on how to use it. The package \texttt{adabmDCA 2.0} is available in different programming languages (C++, Julia, Python) usable on different architectures (single-core and multi-core CPU, GPU) using a common front-end interface. In addition to several learning protocols for dense and sparse generative DCA models, it allows to directly address common downstream tasks like residue-residue contact prediction, mutational-effect prediction, scoring of sequence libraries and generation of artificial sequences for sequence design. It is readily applicable to protein and RNA sequence data.
\end{abstract}


\section{Introduction}
Over the past decade, the increase in sequence data availability has significantly propelled the application of machine learning in biomolecular studies. Generative models, in particular, have emerged as highly effective tools. Even though the field of bioinformatics is moving towards the adoption of larger and larger deep neural network models, notably with the advent of Large Language Models (LLMs), Direct Coupling Analysis (DCA) methods  \cite{cocco_inverse_2018} provide a simple, user-friendly, energy-efficient and interpretable tool that can be successfully adopted for generating functional protein sequences \cite{russ2020evolution}, predicting contacts in the ternary structure \cite{morcos2011direct}, predicting mutational effects in proteins \cite{figliuzzi2016coevolutionary} and unveiling coevolutionary signals in MSAs of homologous protein or RNA families~\cite{cuturello2020assessing}.

This tutorial presents a new version of \texttt{adabmDCA} \cite{muntoni_adabmdca_2021}. The package comes in three different languages: C++ (single-core CPU), Julia (multi-core CPU), and Python (GPU-oriented). They share the same front-end interface from the terminal allowing the user to install and use one of the three equivalent versions based on hardware or software constraints.

We provide three different training routines:
\begin{enumerate}
    \item \textbf{bmDCA}: Trains a fully-connected DCA model \cite{figliuzzi_how_2018};
    \item \textbf{eaDCA}: Trains a DCA model on a sparse coupling network by progressively adding couplings during the training \cite{calvanese_towards_2024};
    \item \textbf{edDCA}: Starts from a trained bmDCA model and iteratively removes the less informative couplings until the target sparsity is reached \cite{barrat-charlaix_sparse_2021}.
\end{enumerate}

Additionally, we provide several routines for sampling and analyzing the generated sequences once a DCA model is trained, for constructing and evaluating - according to a DCA model - a single mutant library from a given wild type, and finally, for computing the pairwise contact scores, in terms of average-product corrected Frobenius norms of the DCA couplings \cite{ekeberg_improved_2013}.

This tutorial is organized as follows. In the first section, we present the theoretical framework of DCA models that is useful for understanding the parameters of the algorithm and the possible issues that one can encounter during the training. We also describe the typical preprocessing pipeline for preparing the data before using them to train the DCA model. In the second section, we describe the different routines and the most important hyperparameters that need to be set. Finally, in the last section, we summarize the main terminal commands for the different routines and we summarize all the input parameters. The reader who is already familiar with DCA modeling of biological sequences can skip the first part of the tutorial and go directly to the third section.

\section{Boltzmann learning of biological models}

\subsection{Structure of the model and Boltzmann learning}

\begin{figure}[t]
    \centering
    \includegraphics[width=0.5\linewidth]{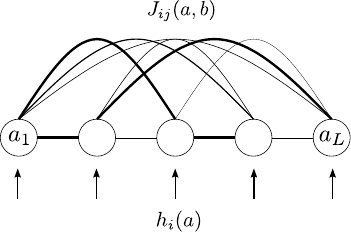}
    \caption{Schematic representation of a DCA model.}
    \label{fig:bmDCA}
\end{figure}

DCA models (Figure~\ref{fig:bmDCA}) are probabilistic generative models that infer a probability distribution over sequence space. Their objective is to assign high probability values to sequences that are statistically similar to the natural sequences used to train the architecture while assigning low probabilities to those that significantly diverge. The input dataset is a Multiple Sequence Alignment (MSA), where each sequence is represented as a $L$-dimensional categorical vector $\pmb a = (a_1, \dots, a_L)$ with $a_i \in \{1, \dots, q\}$, each number representing one of the possible amino acids or nucleotides, or the alignment gap. To simplify the exposition, from here on, we will assume them to be amino acids. The following equation then gives the DCA probability distribution:
\begin{equation}\label{eq:prob DCA}
\begin{split}
    p(\pmb a | \pmb{J}, \pmb{h}, \GG) &= \frac{1}{Z(\pmb{J}, \pmb{h}, \GG)} e^{-E(a_1, \dots, a_L)}\\ & =  \frac{1}{Z(\pmb{J}, \pmb{h}, \GG)}\exp \left( \sum_{(i,a)\in \VV} h_i(a) \delta_{a_i,a} + \sum_{(i,a,j,b)\in \EE} J_{ij}(a, b) \delta_{a_i,a} \delta_{a_j,b}\right).
\end{split}
\end{equation}
In this expression, $Z$ is the normalization constant and $E$ is the DCA \textit{energy function}.
The interaction graph $\GG=(\VV,\EE)$ is 
represented in a {\it one-hot encoding} format: the vertices $\VV$ of the graph are the $L\times q$ combinations of all possible symbols on all possible sites, 
labeled by ${(i,a) \in \{1,\cdots,L\}\times \{1,\cdots,q\}}$, 
and the edges $\EE$ connect two vertices $(i,a)$ and $(j,b)$. 
 The \textit{bias} (or field) $h_i(a)$ corresponding to the amino acid $a$ on the site $i$ 
 is activated by the Kronecker $\delta_{a_i,a}$ if and only if $a_i=a$, 
 and it encodes the conservation signal of the MSA. The \textit{coupling matrix} $J_{ij}(a, b)$ represents the coevolution (or epistatic) signal between pairs of amino acids at different sites, and is activated by the
 $\delta_{a_i,a} \delta_{a_j,b}$ term if $a_i=a$ and $a_j=b$.
 Note that $J_{ii}(a,b)=0$ for all $a,b$ to avoid a redundancy with the $h_i(a)$ terms, and that $J_{ij}(a,b)=J_{ji}(b,a)$ is imposed by the symmetry structure of the model.
The interaction graph $\GG$ can be chosen fully connected as in the bmDCA model, or it can be a sparse graph as in eaDCA and edDCA.

\paragraph{Training the model}
The training consists of adjusting the biases, the coupling matrix, and the interaction graph to maximize the log-likelihood of the model for a given MSA, which can be written as
\begin{align}\label{eq:LL}
    \mathcal{L}(\{\pmb{a}^{(m)}\} | \pmb{J}, \pmb{h},\GG) &= \frac1{M_{\rm eff}}\sum_{m=1}^M w^{(m)} \left[ \sum_{(i,a)\in \VV} h_i(a) \delta_{a_i^{(m)},a} + \sum_{(i,a,j,b)\in \EE} J_{ij}(a, b) \delta_{a_i^{(m)},a} \delta_{a_j^{(m)},b} \right] - \log Z(\pmb{J}, \pmb{h}, \GG) \nonumber \\
    &=  \sum_{(i,a)\in \VV} h_i(a) f_i(a) + \sum_{(i,a,j,b)\in \EE} J_{ij}(a, b) f_{ij}(a,b)  - \log Z(\pmb{J}, \pmb{h}, \GG) \ ,
\end{align}
where $w^{(m)}$ is the weight of the data sequence $m$, with $\sum_{m=1}^M w^{(m)} =M_{\rm eff}$, and
\begin{equation}\label{eq:freqs}
    f_i(a) =\frac1{M_{\rm eff}}\sum_{m=1}^M w^{(m)} \delta_{a_i^{(m)},a} \ ,
    \qquad
        f_{ij}(a,b) = \frac1{M_{\rm eff}}\sum_{m=1}^M w^{(m)}\delta_{a_i^{(m)},a}\delta_{a_j^{(m)},b} \ ,
\end{equation}
are the empirical single-site and two-site frequencies computed from the data.
Roughly speaking, $f_i(a)$ tells us what is the empirical probability of finding the amino acid $a$ in the position $i$ of a training sequence, whereas $f_{ij}(a,b)$ tells us how likely it is to find together the amino acids $a$ and $b$ at positions $i$ and $j$, respectively, in a training sequence. 

For a fixed graph $\GG$, we can maximize the log-likelihood by iteratively updating the parameters of the model in the direction of the gradient of the log-likelihood, meaning

\begin{equation}
    h_i(a) \leftarrow h_i(a) + \gamma \frac{\partial \mathcal{L}(\{\pmb{a}^{(m)}\} | \pmb{J}, \pmb{h}, \GG)}{\partial h_i(a)} \ , \qquad
    J_{i j}(a, b) \leftarrow J_{i j}(a, b) + \gamma \frac{\partial \mathcal{L}(\{\pmb{a}^{(m)}\} | \pmb{J}, \pmb{h}, \GG)}{\partial J_{i j}(a, b)}\ ,
\end{equation}
where $\gamma$ is a small rescaling parameter called \textit{learning rate}. By differentiating the log-likelihood \eqref{eq:LL}, we find the update rule for Boltzmann learning:
\begin{equation}\label{eq:params update}
    h_i(a) \leftarrow h_i(a) + \gamma (f_{i}(a) - p_i(a)) \ , \qquad
    J_{i j}(a, b) \leftarrow J_{i j}(a, b) + \gamma (f_{ij}(a, b) - p_{ij}(a,b)) \ ,
\end{equation}
where $p_i(a) = \langle \delta_{a_i,a}\rangle$ and $p_{ij}(a, b)=\langle \delta_{a_i,a}\delta_{a_j,b}\rangle$ are the one-site and two-site marginals of the model~\eqref{eq:prob DCA}.
Notice that the convergence of the algorithm is reached when $p_i(a) = f_i(a)$ and $p_{ij}(a,b) = f_{ij}(a, b)$.

\paragraph{Monte Carlo estimation of the gradient} 
The difficult part of the algorithm consists of estimating $p_i(a)$ and $p_{ij}(a,b)$, because computing the normalization $Z$ of Eq.~\eqref{eq:prob DCA} is computationally intractable, preventing us from directly computing the probability of any sequence. To tackle this issue, we estimate the first two moments of the distribution through a Monte Carlo simulation. This consists of sampling a certain number of fairly independent sequences from the probability distribution \eqref{eq:prob DCA} and using them to estimate $p_i(a)$ and $p_{ij}(a, b)$ at each learning epoch. There exist several equivalent strategies to deal with it. 
Samples from the model \eqref{eq:prob DCA} can be obtained via Markov Chain Monte Carlo (MCMC) simulations either at equilibrium or out-of-equilibrium, where we start from $N_c$ configurations (we refer to them as \textit{chains}), chosen uniformly at random, from the data or the last configurations of the previous learning epoch, and update them using Gibbs or Metropolis-Hastings sampling steps up to a certain number of MCMC sweeps. 
It has been shown in \cite{muntoni_adabmdca_2021} for Boltzmann machines and, in general, for energy-based models \cite{decelle_equilibrium_2021} that under certain conditions, training the model estimating the marginals from an out-of-equilibrium sampling leads to good enough generative models. Depending on the training set, these models can even be fairly close to what is achieved with perfectly equilibrated training.
For this reason, in \texttt{adabmDCA~2.0}, we implement a persistent contrastive divergence (PCD) scheme, in which the computation of the gradient is done with chains that may be slightly out of equilibrium. 
In particular, chains are \textit{persistent}, i.e. they are initialized at each learning epoch using the last sampled configurations of the previous epoch. This is done because sampling from configurations that are already close to the stationary state of the model at the current training epoch is much more convenient.   Furthermore, the number of sweeps to be performed should be large enough to ensure that the updated chains are close enough to an equilibrium sample of the probability \eqref{eq:prob DCA}. In practice, this is done by fixing the number of sweeps to a convenient value, $k$, that trades off between a reasonable training time and a fair mixing of the chains. 

\paragraph{Convergence criterium}
To decide when to terminate the training, we monitor the two-site connected correlation functions of the data and of the model, which are defined as
\begin{equation}
    C^{\mathrm{data}}_{ij}(a, b) = f_{ij}(a, b) - f_i(a) f_j(b)  \ ,\qquad \qquad
    C^{\mathrm{model}}_{ij}(a, b) = p_{ij}(a, b) - p_i(a) p_j(b) \ .
\end{equation}
When the Pearson correlation coefficient between the two reaches a target value, set by default at 0.95, the training stops.

Once the model is trained, we can generate new sequences by sampling from the probability distribution \eqref{eq:prob DCA}, infer contacts on the tertiary structure by analyzing the coupling matrix, or propose and assess mutations through the DCA energy function $E$ (see Section \ref{sec:app}).
While during the training we keep the number of Monte Carlo sweeps fixed, we underline that
whenever we want to sample new sequences from a model (i.e. after convergence of training) we compute the mixing time as explained in \ref{sec:sampling} and we ensure that sequences are sampled in equilibrium.

\subsection{Training sparse models}
What we have described so far is true for all the DCA models considered in this work, but we have not yet discussed how to adjust the topology of the interaction graph $\GG$. In the most basic implementation, bmDCA, the graph is assumed to be fully connected (every amino acid of any residue is connected with all the rest) and the learning will only tweak the strength of the connections. This results in a coupling matrix with $L (L - 1) q^2 / 2$ independent parameters, where $L$ is the aligned sequence length, $q = 21$ for amino acid and $q=5$ for nucleotide sequences. However, it is well known from the literature that the interaction network of protein families tends to be relatively \textit{sparse}, suggesting that only a few connections should be necessary for reproducing the statistics of biological sequence data. This observation brings us to devising a training routine that produces a DCA model capable of reproducing the one and two-body statistics of the data with a minimal amount of couplings.

To achieve this, we implemented two different routines: eaDCA that promotes initially inactive (i.e. zero) coupling parameters to active ones starting from a profile model, and edDCA, which iteratively prunes active but negligible coupling parameters starting from a dense fully-connected DCA model.

\begin{figure}[t]
    \centering
    \includegraphics[width=\linewidth]{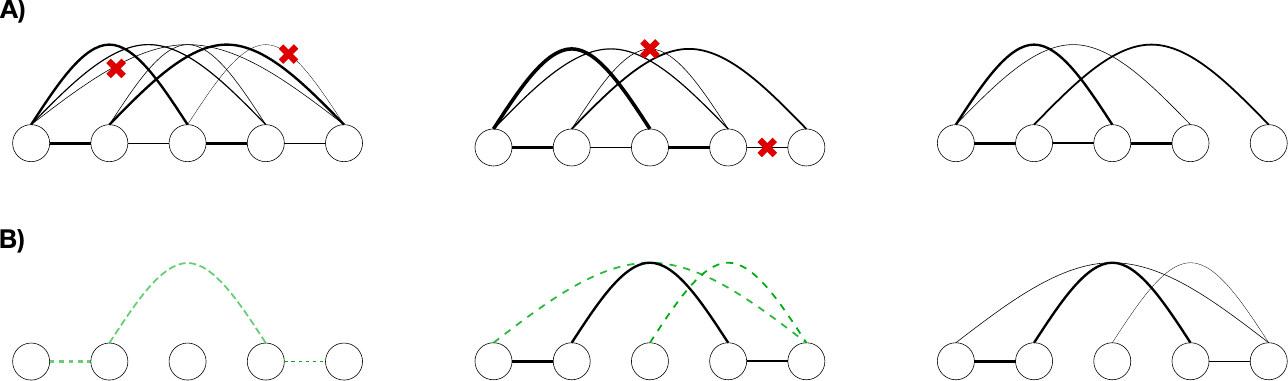}
    \caption{Schematic representation of the sparse model training. A) edDCA, the sparsification is obtained by progressively pruning contacts from an initial fully connected model. B) eaDCA, the couplings are progressively added during the training.}
    \label{fig:sparseDCA}
\end{figure}

\paragraph{Element activation DCA}
In eaDCA (Figure~\ref{fig:sparseDCA}-B), we start from an empty interaction graph $\EE = \oslash$, meaning that no connection is present. Each training step is divided into two different moments: first, we update the graph, and then we bring the model to convergence once the graph is fixed (a similar pipeline has been proposed in~\cite{calvanese_towards_2024}). 

To update the graph, we first estimate $p_{ij}(a, b)$ for the current model. Then, we decide how many couplings we want to target as the fraction \texttt{factivate} of the number of currently inactive couplings and, from all the possible quadruplets of indices $(i, j, a, b)$, we select those for which $p_{ij}(a, b)$ is ``the most distant'' (according to the criterion introduced in~\cite{calvanese_towards_2024}) from the target statistics $f_{ij}(a,b)$. We then activate the couplings corresponding to these quadruplets, obtaining a new graph $\EE' \supseteq \EE$. Notice that some of the selected couplings might be already active, so the model partially auto-regulates the number of new couplings that have to be activated at each step.

\paragraph{Element decimation DCA}
In edDCA (Figure~\ref{fig:sparseDCA}-A), we start from a previously trained bmDCA model and its fully connected graph $\GG$. We then apply the decimation algorithm, in which we prune connections from the edges $\EE$ until a target density of the graph is reached, where the density is defined as the ratio between the number of active couplings and the number of couplings of the fully connected model. Similarly to eaDCA, each iteration consists of two separate moments: graph updating and active parameter updating.

To update the graph, we remove the fraction \texttt{drate} of active couplings that, once removed, produce the smallest perturbation on the probability distribution at the current epoch. In particular, for each active coupling, one computes the symmetric Kullback-Leibler distances between the current model and a perturbed one, without that target element. One then removes the \texttt{drate} elements which exhibit the smallest distances (see~\cite{barrat-charlaix_sparse_2021} for further details).

\paragraph{Parameter updates in between decimations/activations}
In both procedures, to bring the model to convergence on the graph, we perform a certain number of parameter updates in  between each step of edge activation or decimation, using the update rule \eqref{eq:params update}. Between two subsequent parameter updates, $k$ sweeps are performed to update the Markov chains.

In the case of element activation we perform a
fixed number of parameter updates, 
specified by the input parameter \texttt{gsteps}.
Instead, when pruning the graph, we keep updating the parameters with the rule \eqref{eq:params update} until the Pearson correlation coefficient reaches a target value.

\subsection{Input data and pre-processing}
\texttt{adabmDCA~2.0} takes as input a multiple sequence alignment (\href{https://en.wikipedia.org/wiki/Multiple_sequence_alignment}{MSA}) of aligned amino acid or nucleotide sequences, usually forming a protein or RNA family. DCA implementations require the data to be saved in FASTA format~\cite{pearson_improved_1988}. 

\texttt{adabmDCA~2.0} implements the three default alphabets shown in table~\ref{tab:alphabets}, but the user can specify an ad-hoc alphabet as far as it is compatible with the input MSA.

\begin{table}[ht!]
    \centering
    \begin{tabular}{c|c}
    \textbf{protein}     &  \texttt{-, A, C, D, E, F, G, H, I, K, L, M, N, P, Q, R, S, T, V, W, Y} \\
    \textbf{rna}     & \texttt{-, A, C, G, U} \\
    \textbf{dna}     & \texttt{-, A, C, G, T}
    \end{tabular}
    \caption{Default alphabets implemented in \texttt{adabmDCA~2.0}, where ``-'' indicates the alignment gap.}
    \label{tab:alphabets}
\end{table}

\begin{figure}[!ht]
    \centering
    \includegraphics[width=0.9\textwidth]{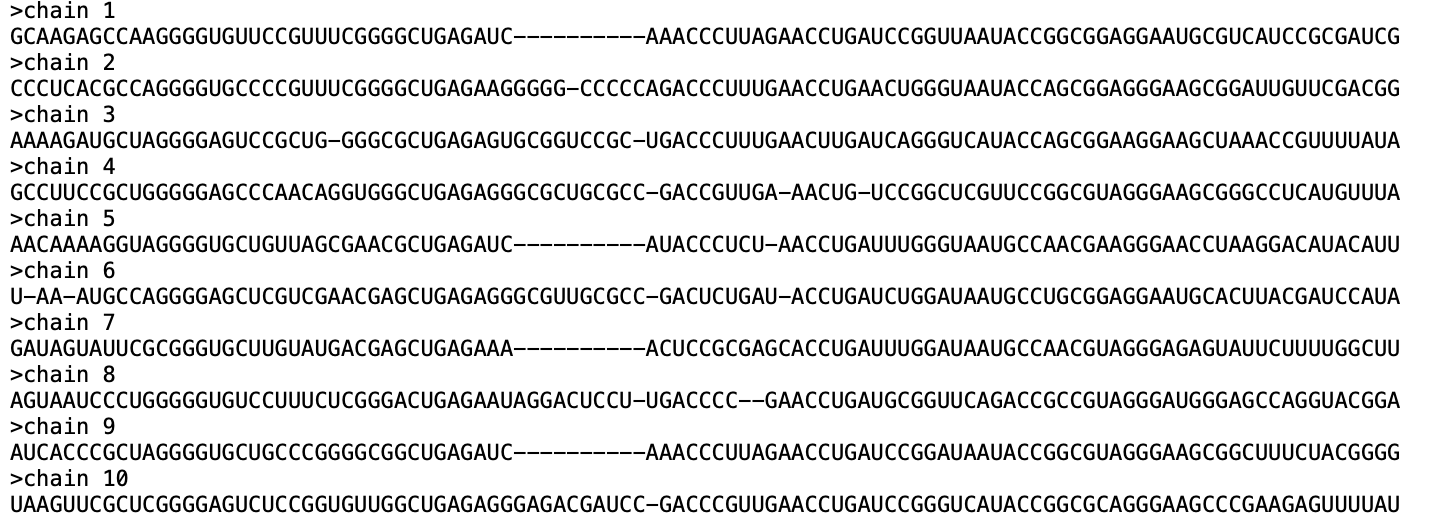}
    \caption{Example of RNA sequences formatted in FASTA format.}
    \label{fig:example fasta}
\end{figure}

An example of a FASTA file format is shown in Figure~\ref{fig:example fasta}. In particular, \texttt{adabmDCA 2.0} correctly handles FASTA files in which line breaks within a sequence are present.

\subsubsection{Preprocessing}

\paragraph{Preprocessing pipeline}


The \texttt{adabmDCA 2.0} code applies the following preprocessing pipeline to the input MSA:
\begin{itemize}
    \item[p.1] Remove the sequences having some tokens not included in the chosen alphabet;
    \item[p.2] If needed, compute the importance weights for the sequences in the MSA;
    \item[p.3] Apply a pseudocount to compute the MSA statistics.
\end{itemize}
Their precise implementation is described in the following.

\paragraph{Computing the importance weights}\label{sec:importance weights}
The sequence weights are computed to mitigate as much as possible the systematic biases in the data, such as correlations due to the phylogeny or over-representation of some regions of the sequence space because of a sequencing bias.
Given an MSA of $M$ sequences, to compute the importance weight of each sequence $\pmb a^{(m)}$, $m=1, \dots, M$, we consider $N^{(m)}$ as the number of sequences in the dataset having sequence identity from $\pmb a^{(m)}$ greater or equal to $0.8 \cdot L$ (this threshold can be tuned by the user). Then, the importance  weight of $\pmb{a}^{(m)}$ will be
\begin{equation}\label{eq:weights}
    w^{(m)} = \frac{1}{N^{(m)}} \ .
\end{equation}
This reweighting allows us to give less importance to sequences found in very densely populated regions of the sequence space while enhancing the importance of isolated sequences. Alternatively, the user can decide to provide an external file containing the weights, or to ask the program to assign uniform weights across the sequences in the MSA.

\paragraph{Pseudocount and reweighted statistics}
DCA models are trained to reproduce the one and two-site frequencies of the empirical data. To compute these, we introduce in the computation of the empirical statistics a small parameter $\alpha$, called pseudocount, that allows us to deal with unobserved (pairs of) symbols in one (or two) column(s) of the MSA. The one and two-site frequencies are given by

\begin{align}
    \label{eq:freq1}
    f_i(a) &= (1 - \alpha) f^{\mathrm{data}}_i(a) + \frac{\alpha}{q} \ , \\
    \label{eq:freq2}
    f_{ij}(a, b) &= (1 - \alpha) f^{\mathrm{data}}_{ij}(a, b) + \frac{\alpha}{q^2} \ .
\end{align}
where $f_i^{\mathrm{data}}(a)$ and $f_{ij}^{\mathrm{data}}(a, b)$ are computed from the MSA as in Eq.~\eqref{eq:freqs}.

\section{Implementation}
All the software implementations that we propose (Python, Julia, and C++) offer the same interface from the terminal through the \texttt{adabmDCA} command.
The complete list of training options can be listed through the command

\begin{lstlisting}[language=bash, basicstyle=\footnotesize]
  $ adabmDCA train -h
\end{lstlisting}
The standard command for starting the training of a DCA model is

\begin{lstlisting}[language=bash, basicstyle=\footnotesize]
  $ adabmDCA train -m <model> -d <fasta_file> -o <output_folder> -l <label>
\end{lstlisting}
where
\begin{itemize}
    \item \texttt{<model>}$~\in \{\text{bmDCA},~\text{eaDCA}, ~\text{edDCA}\}$ selects the training routine. By default, the fully connected bmDCA algorithm is used. edDCA can follow two different routines: either it decimates a pre-trained bmDCA model, or it first trains a bmDCA model and then decimates it. The corresponding commands are shown below (see section \ref{sec: decDCA}); 
    \item \texttt{<fasta\_file>} is the FASTA file, with the complete path, containing the training MSA; 
    \item \texttt{<output\_folder>} is the path to a (existing or not) folder where to store the output files;
    \item \texttt{<label>} is an optional argument. If provided, it will label the output files. This is helpful when running the algorithm multiple times in the same output folder. 
\end{itemize}

\begin{figure}[t]
    \centering
    \includegraphics[width=0.5\linewidth]{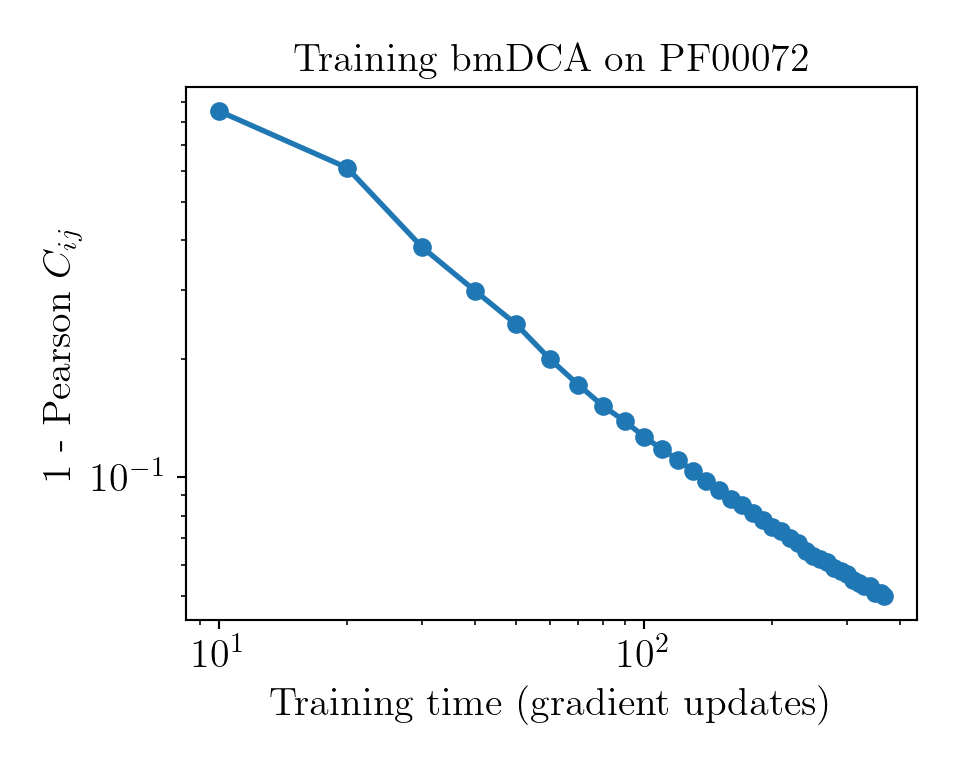}
    \caption{One minus the Pearson correlation coefficient between the data and chains correlation matrices as a function of the training time for the protein family PF00072. The curve is well approximated by a power law decay.}
    \label{fig:Pearson vs epochs}
\end{figure}

Once started, the training will continue until the Pearson correlation coefficient between the two-point connected correlations of the model and the empirical ones obtained from the data reaches a modifiable target value (set by default at \texttt{target = 0.95}). In figure~\ref{fig:Pearson vs epochs} we show in a log-log plot the evolution of (one minus) the Pearson correlation coefficient as a function of the training time. The long-time evolution of the training is well approximated by a straight line, meaning that the Pearson growth follows a power law. As such, one should keep in mind that the training procedure converges very rapidly at the beginning (say, it reaches Pearson values of $\sim 0.9$ in about 100 iterations in figure~\ref{fig:Pearson vs epochs}), while reaching higher values of the Pearson requires comparatively more algorithmic time. Therefore, if needed, it is possible to fit a coarse model in a very short amount of time by just lowering the \texttt{target} value somewhere around~0.9.

\paragraph{Output files}
By default the training algorithm outputs three kinds of text files:


\begin{itemize}
    \item \texttt{<label>\_params.dat}: 
    file containing the parameters of the model saved in this format:
    \begin{itemize}
        \item Lines starting with \texttt{J} represent entries of the coupling matrix, followed by the two interacting positions in the sequence and the two amino acids or nucleotides involved.
        \item Lines starting with \texttt{h} represent the bias, followed by a number and a letter indicating the position and the amino acid or nucleotide subject to the bias.
    \end{itemize}
    Note that inactive, i.e. zero couplings are not included in the file.

    \begin{figure}[!ht]
        \centering
        \includegraphics[width=0.35\linewidth]{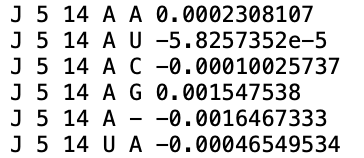}
        \\
        \includegraphics[width=0.35\linewidth]{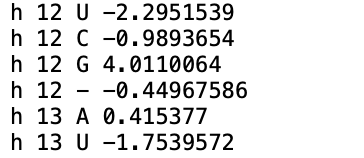}
        \caption{Format for saving the coupling matrix \( J \) and biases \( h \) for an RNA domain.
        }
        \label{fig:format_model}
    \end{figure}
    \item \texttt{<label>\_chains.fasta}: FASTA file containing the sequences corresponding to the last state of the Markov chains used during the learning;
    \item \texttt{<label>\_adabmDCA.log}: .log file containing information collected during the training procedure.
\end{itemize}
During the training, the output files containing the parameters and the chains are overwritten, and the log file is updated, every $50$ gradient updates (epochs) for \texttt{bmDCA}, and every 10 graph updates for \texttt{eaDCA} and \texttt{edDCA}.

\paragraph{Restore an interrupted training}
It is possible to start the training by initializing the parameters of the model and the chains at a given checkpoint. To do so, two arguments specifying the path of the parameters and the chains are needed:
\begin{lstlisting}[language=bash, basicstyle=\footnotesize]
  $ adabmDCA train [...] -p <file_params> -c <file_chains>
\end{lstlisting}

\paragraph{Importance weights}
It is possible to provide the algorithm with a pre-computed list of importance weights to be assigned to the sequences (see section \ref{sec:importance weights}) by giving the path to the text file to the argument \texttt{-w}. If this argument is not provided, the algorithm will automatically compute the weights using Eq.~\eqref{eq:weights} and it will store them into the folder \texttt{<output\_folder>} as \texttt{<label>\_weights.dat}. The default sequence identity threshold used for computing the weights is set to 0.8, but this value can be modified using the argument \texttt{--clustering\_seqid}. It is also possible to ask the routine to use uniform weights for all sequences by means of the flag \texttt{--no\_reweighting}.


\paragraph{Choosing the alphabet}
By default, the algorithm will assume that the input MSA belongs to a protein family, and it will use the preset alphabet defined in Table \ref{tab:alphabets} (by default: \texttt{--alphabet protein}). If the input data comes from RNA or DNA sequences, it has to be specified by passing respectively \texttt{rna} or \texttt{dna} to the \texttt{--alphabet} argument. There is also the possibility of passing a user-defined alphabet, provided that all the tokens match with those that are found in the input MSA. This can be useful if one wants to use a different order than the default one for the tokens, or in the eventuality that one wants to handle additional symbols present in the alignment.
This is done using \texttt{--alphabet ABCD-} if for example the alphabet contains the symbols \texttt{A, B, C, D, -}.

\subsection{eaDCA}
To train an eaDCA model, we just have to specify \texttt{--model eaDCA}. Two more hyperparameters can be changed:
\begin{itemize}
    \item \texttt{--factivate}: The fraction of inactive couplings that are selected for the activation at each update of the graph. By default, it is set to 0.001.
    \item \texttt{--gsteps}: The number of parameter updates to be performed on the given graph. By default, it is set to 10.
\end{itemize}

For this routine, the number of sweeps for updating the chains can be typically reduced to 5, since only a fraction of all the possible couplings have to be updated at each iteration.

\subsection{edDCA}\label{sec: decDCA}

To launch a decimation with default hyperparameters, use the command:

\begin{lstlisting}[language=bash, basicstyle=\footnotesize]
  $ adabmDCA train -m edDCA -d <fasta_file> -p <file_params> -c <file_chains>
\end{lstlisting}
where \texttt{<file\_params>} and \texttt{<file\_chains>} are, respectively, the file names of the parameters and the chains (including the path) of a previously trained bmDCA model. The edDCA can perform two routines as described above. In the first routine, it uses a pre-trained bmDCA model and its associated chains provided through the parameters \texttt{<file\_params>} and \texttt{<file\_chains>}. The routine makes sure everything has converged before starting the decimation of couplings. It repeats this process for up to 10,000 iterations if needed. If these parameters are not supplied, the second routine initializes the model and chains randomly, trains the bmDCA model to meet convergence criteria, and then starts the decimation process as described.
Some important parameters that can be changed are:
\begin{itemize}
    \item \texttt{--gsteps}: The number of parameter updates to be performed at each step of the convergence process on the given graph. By default, it is set to 10.
    \item \texttt{--drate}: Fraction of active couplings to be pruned at each graph update. By default, it is set to 0.01.
    \item \texttt{--density}: Density of the graph that has to be reached after the decimation. The density is defined as the ratio between the number of active couplings and the number of couplings of the fully connected graph. By default, it is set to 0.02.
    \item \texttt{--target}: The Pearson correlation coefficient to be reached to assume the model to have converged. By default, it is set to 0.95. 
\end{itemize}

\subsection{How to choose the hyperparameters?}
The default values for the hyperparameters are chosen to be a good compromise between having a relatively short training time and a good quality of the learned model for most of the \textit{typical} input MSA, where for \textit{typical} we mean a clean MSA with only a few gaps for each sequence and a not too structured dataset (not too clustered in subfamilies) that could create some ergodicity problems during the training. It may happen, though, that for some datasets some adjustments of the hyperparameters are needed to get a properly trained model. The most important ones are:

\paragraph{Learning rate}
By default, the learning rate is set to 0.05, which is a reasonable value in most cases. If the resampling of the model is bad (very long thermalization time or mode collapse), one may try to decrease the learning rate through the argument \texttt{--lr} to some smaller value (e.g. 0.01 or 0.005).

\paragraph{Number of Markov Chains}
By default, the number of Markov chains is set to 10000, which works well in most cases. Using fewer chains reduces the memory required to train the model, but it may also lead to a longer algorithm convergence time.
To change the number of chains, we can use the argument \texttt{--nchains}.

\paragraph{Number of Monte Carlo steps}
The argument \texttt{--nsweeps} defines the number of Monte Carlo chain updates (sweeps) between one gradient update and the following. A single \textit{sweep} is obtained once we propose a mutation for all the residues of the sequence. By default, this parameter is set to 10, which is a good choice for easily tractable MSAs. The higher this number is chosen, the better the quality of the training will be, because in this way we allow the Markov chains to decorrelate more to the previous configuration. However, this parameter heavily impacts the training time, so we recommend choosing it in the interval 10 - 50.

\paragraph{Regularization}
Another parameter that can be adjusted if the model does not resample correctly is the pseudocount, $\alpha$, which can be changed using the key \texttt{--pseudocount}. The pseudocount is a regularization term that introduces a flat prior on the frequency profiles, modifying the frequencies as in equations \eqref{eq:freq1} and \eqref{eq:freq2}.
If $\alpha = 1$ we impose an equal probability to all the amino acids to be found in the residue at position $i$, while if $\alpha = 0$ we just use the empirical frequencies of the data. By default, the pseudocount is set as the inverse of the effective number of sequences, $1 / M_{\mathrm{eff}}$, where 
\begin{equation}
    M_{\mathrm{eff}} = \sum_{m=1}^M w^{(m)}.
\end{equation}

\section{Applications \label{sec:app}}
\subsection{Generate sequences}\label{sec:sampling}

\begin{figure}[!ht]
    \centering
    \includegraphics[width=\linewidth]{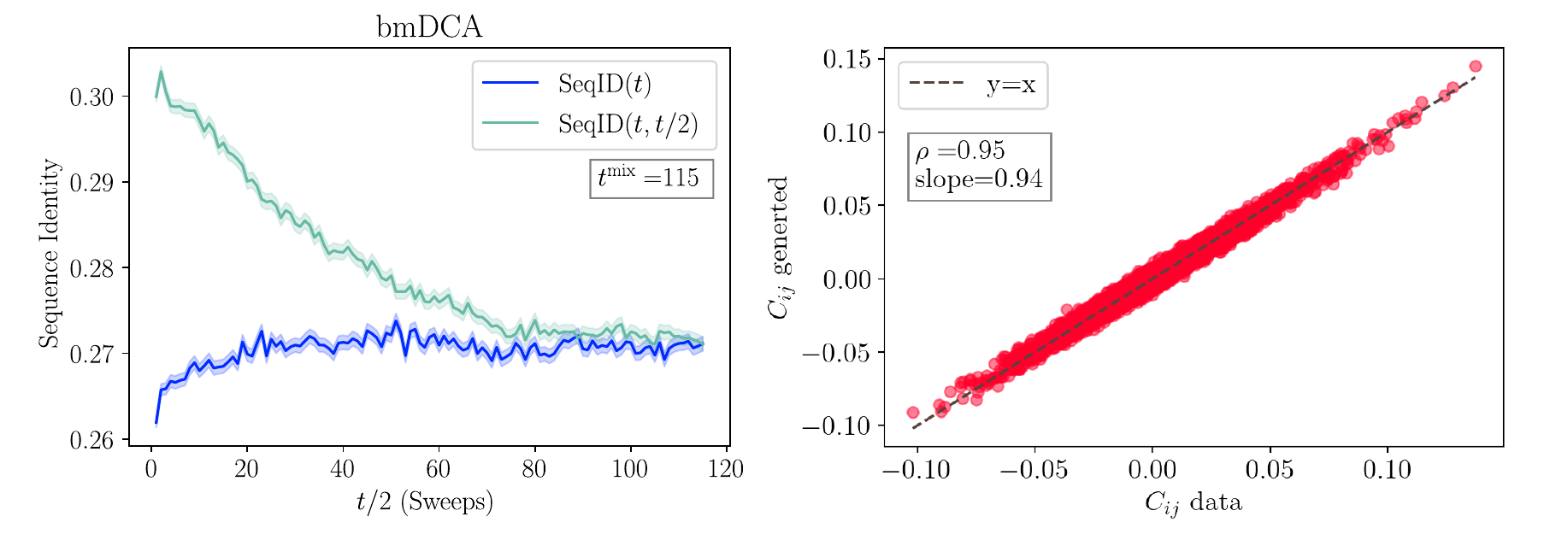}
    \caption{Analysis of a bmDCA model. \textbf{Left}: measuring the mixing time of the model using $10^4$ chains. The curves represent the average overlap among randomly initialized samples (dark blue) and the one among the same sequences between times $t$ and $t/2$ (light blue). Shaded areas represent the error of the mean. When the two curves merge, we can assume that the chains at time $t$ forgot the memory of the chains at time $t/2$. This point gives us an estimate of the mixing time of the model, $t^{\mathrm{mix}}$.  Notice that the times start from 1, so the starting conditions are not shown. \textbf{Right}: Scatter plot of the entries of the covariance matrix of the data versus that of the generated samples.}
    \label{fig:mixing time}
\end{figure}

Once we have a trained model, we can use it to generate new sequences. This can be done using the command:
\begin{lstlisting}[language=bash, basicstyle=\footnotesize]
  $ adabmDCA sample -p <path_params> -d <fasta_file> -o <output_folder> --ngen <num_gen>
\end{lstlisting}
where \texttt{output\_folder} is the directory where to save the data and \texttt{num\_gen} is the number of sequences to be generated. The routine will first compute the mixing time ($t^{\mathrm{mix}}$) of the model by running a simulation starting from the sequences of the input MSA. After that, it will randomly initialize \texttt{num\_gen} Markov chains and run for \texttt{nmix}~$\times~t^{\mathrm{mix}}$ sweeps to ensure that the model equilibrates. It will save in the output directory a FASTA file containing the sequences sampled with the model and a text file containing the records used to determine the convergence of the algorithm. Figure \ref{fig:mixing time}
-Right shows the comparison between the entries of the covariance matrices obtained from the data and from the generated sequences. The Pearson correlation coefficient is the same used as a target for the training and the slope is close to 1, meaning that the model is able to correctly recover the two-sites statistics of the data MSA.

\paragraph{Convergence criterion}
To determine the convergence of the Monte Carlo simulation, the following strategy is used. We extract \texttt{nmeasure}~$=N$ sequences from the data MSA according to their statistical weight and we make a copy of them, which are used to initialize a set of $N$ chains. We then compare two sets of chains. The first set represents the chains simulated up to time $t$, and we denote them as $\pmb{A}(t) = \{\pmb{a}_1(t), \dots, \pmb{a}_{N}(t)\}$, while the sequences of the second set are the chains simulated until time $t/2$, and we call them $\pmb{A}(t/2) = \{\pmb{a}_1(t/2), \dots, \pmb{a}_{N}(t/2)\}$.
With some abuse of notation, we define the \textit{intrachain correlation} and \textit{autocorrelation} as
\begin{equation}
    \mathrm{SeqID}(t) = \frac{1}{N} \sum_{i=1}^{N} \mathrm{SeqID}(\pmb{a}_i(t), \pmb{a}_{\sigma(i)}(t)) \ , \qquad \mathrm{SeqID}(t, t/2) = \frac{1}{N} \sum_{i=1}^{N} \mathrm{SeqID}(\pmb{a}_i(t), \pmb{a}_i(t/2)) \ ,
\end{equation}
where $\sigma(i)$ is a random permutation of the index $i$ and
\begin{equation}
    \mathrm{SeqID}(\pmb{a}, \pmb{b}) = \frac{1}{L}\sum_{i=1}^L \delta_{a_i, b_i} \in [0, 1]
\end{equation}
is the normalized sequence identity (or overlap) between the sequences $\pmb{a}$ and $\pmb{b}$. By construction, at the initialization we have $\mathrm{SeqID}(t, t/2) = 1$ and $\mathrm{SeqID}(t)$ somewhat close to the average sequence identity of the MSA.  The convergence is obtained when chains are \textit{mixed}, meaning that the system has completely forgotten the initial configuration. This requirement is satisfied when the statistics of a set of independent chains is the same as the one between the initialization and the evolved chains, meaning $\mathrm{SeqID}(t) \cong \mathrm{SeqID}(t, t/2)$. The point at which the two curves merge is called \textit{mixing time}, and we denote it as $t^{\mathrm{mix}}$. After reaching the mixing time of the model, the algorithm will initialize \texttt{ngen} chains at random. It will run a sampling for other \texttt{nmix}~$\times~t^{\mathrm{mix}}$ steps to guarantee complete thermalization, with \texttt{nmix=2} by default. Together with the generated sequences, the script will output a text file containing the records of $\mathrm{SeqID}(t)$ and $\mathrm{SeqID}(t, t/2)$ and their standard deviations  (figure \ref{fig:mixing time}-Left).

\subsection{Contact prediction}
One of the principal applications of the DCA models has been that of predicting a tertiary structure of a protein or RNA domain. In particular, with each pair of sites $i$ and $j$ in the MSA, \texttt{adabmDCA 2.0} computes a contact score that quantifies how likely the two associated positions in the chains are in contact in the three-dimensional structure.
Formally, it corresponds to the average-product corrected (APC) Frobenius norms of the coupling matrices~\cite{ekeberg_improved_2013}, i.e.
\begin{equation}
F_{i,j}^{\rm APC} = F_{i,j} - \frac{\sum_{k} F_{i,k} \sum_{k} F_{k,j}}{\sum_{kl} F_{k,l}} \ , \qquad F_{i,j} = \sqrt{\sum_{a,b \neq `-`} J_{i,j}\left(a, b \right)^{2}} \ .
\label{eq:frob}
\end{equation}
Note that the coupling parameters are usually transformed in a zero-sum gauge before computing the scores, and the gap symbol should be neglected while computing the sum in Eq. \ref{eq:frob}~\cite{feinauer_improving_2014}. 
The scores for all site pairs are provided in the output folder in a separate file called \texttt{<label>\_frobenius.txt}. The first two columns indicate the site indices and the third one contains the associated APC Frobenius norm. The command for computing the matrix of Frobenius norms is
\begin{lstlisting}[language=bash, basicstyle=\footnotesize]
  $ adabmDCA contacts -p <file_params>  -o <output_folder>
\end{lstlisting}

\subsection{Scoring a sequence set}
At convergence, users can score a set of input sequences according to a trained DCA model by using the command line
\begin{lstlisting}[language=bash, basicstyle=\footnotesize]
  $ adabmDCA energies -d <fasta_file>  -p <file_params>  -o <output_folder>
\end{lstlisting}
\texttt{adabmDCA~2.0} will produce a new FASTA file in the output folder \texttt{<output\_folder>} where the input sequences in \texttt{<fasta\_file>} have an additional field in the name that account for the DCA energy function computed according to the model in \texttt{<file\_params>}. Note that low energies correspond to good sequences.

\subsection{Single mutant library}
Another possible application exploits the sequence-fitness score computable according to the energy function $E$ in Eq.~\eqref{eq:prob DCA}. This routine provides a single-mutant library for a given wild type to possibly guide Deep Mutational Scanning (DMS) experiments. In particular, \texttt{adabmDCA 2.0} allows one to predict the fitness reduction (increase) in terms of $\Delta E = E\left(\rm mutant \right) - E\left(\rm wildtype \right)$ for positive (negative) value of $\Delta E$, respectively~\cite{hopf_mutation_2017}.
To produce a FASTA file containing all weighted single-mutants one has to run
\begin{lstlisting}[language=bash, basicstyle=\footnotesize]
  $ adabmDCA DMS -d <WT> -p <file_params> -o <output_folder>
\end{lstlisting}
where \texttt{<WT>} is the name of the FASTA file containing the wild type sequence, \texttt{<file\_params>} is a model file in a format compatible with \texttt{adabmDCA~2.0} output, and \texttt{<output\_folder>} corresponds to the folder that will contain the library file. The sequences in the output FASTA file are named after the introduced mutation and the corresponding $\Delta E$; for instance, \texttt{>G27A~|~DCAscore:~-0.6} denotes that position \texttt{27} has been changed from \texttt{G} to \texttt{A} and $\Delta E = -0.6$. 

\section{Source Code}
The code and the documentation for the \texttt{adabmDCA 2.0} package can be found at \url{https://github.com/spqb/adabmDCA}. We also prepared a \href{https://colab.research.google.com/drive/1l5e1W8pk4cB92JAlBElLzpkEk6Hdjk7B?usp=sharing}{Colab} notebook with a simple tutorial for training and sampling a \texttt{bmDCA} model.

\section{Quicklist}

\subsection{Commands}

\begin{itemize}
    \item Train a bmDCA model with default arguments:
    \begin{lstlisting}[language=bash, basicstyle=\scriptsize]
  $ adabmDCA train -d <fasta_file> -o <output_folder>
    \end{lstlisting}

    \item Restore the training of a bmDCA model:
    \begin{lstlisting}[language=bash, basicstyle=\scriptsize]
  $ adabmDCA train -d <fasta_file> -o <output_folder> -p <file_params> -c <file_chains>
    \end{lstlisting}

    \item Train an eaDCA model with default arguments:
    \begin{lstlisting}[language=bash, basicstyle=\scriptsize]
  $ adabmDCA train -m eaDCA -d <fasta_file> -o <output_folder> --nsweeps 5
    \end{lstlisting}

    \item Restore the training of an eaDCA model:
    \begin{lstlisting}[language=bash, basicstyle=\scriptsize]
  $ adabmDCA train -m eaDCA -d <fasta_file> -o <output_folder> -p <file_params> -c <file_chains>
    \end{lstlisting}

    \item Decimate a bmDCA model at 2\% of density:
    \begin{lstlisting}[language=bash, basicstyle=\scriptsize]
  $ adabmDCA train -m edDCA -d <fasta_file> -p <file_params> -c <file_chains>
    \end{lstlisting}

    \item Train and decimate a bmDCA model at 2\% of density:
    \begin{lstlisting}[language=bash, basicstyle=\scriptsize]
  $ adabmDCA train -m edDCA -d <fasta_file> 
    \end{lstlisting}

    \item Sample from a previously trained DCA model:
    \begin{lstlisting}[language=bash, basicstyle=\scriptsize]
  $ adabmDCA sample -p <file_params> -d <fasta_file> -o <output_folder> --ngen <num_gen>
    \end{lstlisting}

    \item Scoring a sequence set:
    \begin{lstlisting}[language=bash, basicstyle=\scriptsize]
  $ adabmDCA  energies -d <fasta_file>  -p <file_params>  -o <output_folder>
\end{lstlisting}

    \item Generating a single mutant library starting from a wild type:
    \begin{lstlisting}[language=bash, basicstyle=\scriptsize]
  $ adabmDCA DMS -d <WT> -p <file_params> -o <output_folder>
\end{lstlisting}
\item Computing the matrix of Frobenius norms for the contact prediction:
\begin{lstlisting}[language=bash, basicstyle=\scriptsize]
  $ adabmDCA contacts -p <file_params> -o <output_folder>
\end{lstlisting}

\end{itemize}

\subsection{Script arguments}

\begin{table}[h]
    \centering
    \begin{tabular}{llp{8cm}}
        \toprule
        \textbf{Command} & \textbf{Default value} & \textbf{Description} \\
        \midrule
        \texttt{-d, --data} & \centering N/A & Filename of the dataset to be used for training the model. \\
        \texttt{-o, --output} & \centering DCA\_model & Path to the folder where to save the model. \\
        \texttt{-m, --model} & \centering bmDCA & Type of model to be trained. Possible options are \texttt{bmDCA}, \texttt{eaDCA} and \texttt{edDCA}. \\
        \texttt{-w, --weights} & \centering None & Path to the file containing the weights of the sequences. If \texttt{None}, the weights are computed automatically. \\
        \texttt{--clustering\_seqid} & \centering 0.8 & Sequence identity threshold to be used for computing the sequence weights \\
        \texttt{--no\_reweighting} & \centering N/A & If this flag is used, the routine assigns uniform weights to the sequences \\
        \texttt{-p, --path\_params} & \centering None & Path to the file containing the model's parameters. Required for restoring the training. \\
        \texttt{-c, --path\_chains} & \centering None & Path to the FASTA file containing the model's chains. Required for restoring the training. \\

       \texttt{-l, --label} & \centering None & A label to identify different algorithm runs. It prefixes the output files with this label.\\
                
        \texttt{--alphabet} & \centering protein & Type of encoding for the sequences. Choose among \texttt{protein}, \texttt{rna}, \texttt{dna} or a user-defined string of tokens. \\
        \texttt{--lr} & \centering 0.01 & Learning rate. \\
        \texttt{--nsweeps} & \centering 10 & Number of sweeps for each gradient estimation. \\
        \texttt{--sampler} & \centering gibbs & Sampling method to be used. Possible options are \texttt{gibbs} and \texttt{metropolis}. \\
        \texttt{--nchains} & \centering 10000 & Number of Markov chains to run in parallel. \\
        \texttt{--target} & \centering 0.95 & Pearson correlation coefficient on the two-sites statistics to be reached. \\
        \texttt{--nepochs} & \centering 50000 & Maximum number of epochs allowed. \\
        \texttt{--pseudocount} & \centering None & Pseudo count for the single and two-sites statistics. Acts as a regularization. If \texttt{None}, it is set to $1/M_{\mathrm{eff}}$. \\
        \texttt{--seed} & \centering 0 & Random seed.\\
        \texttt{--nthreads}\footnotemark[1] & \centering 1 & Number of threads used in the Julia multithreads version. \\
        \texttt{--device}\footnotemark[1] & \centering cuda & Device to be used between cuda (GPU) and CPU. Used in the Python version. \\
        \texttt{--dtype}\footnotemark[1] & \centering float32 & Data type to be used between float32 and float64. Used in the Python version. \\
        \midrule
        \midrule
        \multicolumn{3}{c}{\textbf{eaDCA options}} \\
        \midrule
        \midrule
        \texttt{--gsteps} & \centering 10 & Number of gradient updates to be performed on a given graph. \\
        \texttt{--factivate} & \centering 0.001 & Fraction of inactive couplings to try to activate at each graph update. \\
        \midrule
        \midrule
        \multicolumn{3}{c}{\textbf{edDCA options}} \\
        \midrule
        \midrule
          \\
        \texttt{--gsteps} & \centering 10 & The number of gradient updates applied at each step of the graph convergence process. \\
        \texttt{--density} & \centering 0.02 & Target density to be reached. \\
        \texttt{--drate} & \centering 0.01 & Fraction of remaining couplings to be pruned at each decimation step. \\
        \bottomrule
    \end{tabular}
    \caption{Command line arguments for training bmDCA, eaDCA or edDCA.}
    \label{table:arguments bmDCA}
\end{table}

\begin{table}[h]
    \centering
    \begin{tabular}{llp{8cm}}
        \toprule
        \textbf{Command} & \textbf{Default value} & \textbf{Description} \\
        \midrule
        \texttt{-p, --path\_params} & \centering N/A & Path to the file containing the parameters of the DCA model to sample from. \\
        \texttt{-d, --data} & \centering N/A & Filename of the dataset MSA. \\
        \texttt{-o, --output} & \centering N/A & Path to the folder where to save the output. \\
        \texttt{--ngen} & \centering None & Number of samples to generate.\\
        \texttt{-l, --label} & \centering None & A label to identify different algorithm runs. It prefixes the output files with this label.\\
        \texttt{-w, --weights} & \centering None & Path to the file containing the weights of the sequences. If \texttt{None}, the weights are computed automatically. \\
        \texttt{--clustering\_seqid} & \centering 0.8 & Sequence identity threshold to be used for computing the sequence weights \\
        \texttt{--no\_reweighting} & \centering N/A & If this flag is used, the routine assigns uniform weights to the sequences \\
        \texttt{--nmeasure} & \centering 10000 & Number of data sequences to use for computing the mixing time. The value min(\texttt{nmeasure}, len(data)) is taken.\\
        \texttt{--nmix} & \centering 2 & Number of mixing times used to generate 'ngen' sequences starting from random. \\
        \texttt{--max\_nsweeps} & \centering 10000 & Maximum number of sweeps allowed. \\
        \texttt{--alphabet} & \centering protein & Type of encoding for the sequences. Choose among \texttt{protein}, \texttt{rna}, \texttt{dna} or a user-defined string of tokens. \\
        \texttt{--sampler} & \centering gibbs & Sampling method to be used. Possible options are \texttt{gibbs} and \texttt{metropolis}. \\
        \texttt{--beta} & \centering 1.0 & Inverse temperature to be used for the sampling. \\
        \texttt{--pseudocount} & \centering None & Pseudo count for the single and two-sites statistics. Acts as a regularization. If \texttt{None}, it is set to $1/M_{\mathrm{eff}}$. \\
        \texttt{--device}\footnotemark[1] & \centering cuda & Device to be used between cuda (GPU) and CPU. Used in the Python version. \\
        \texttt{--dtype}\footnotemark[1] & \centering float32 & Data type to be used between float32 and float64. Used in the Python version. \\
        \bottomrule
    \end{tabular}
    \caption{Command line arguments for sampling from a DCA model.}
    \label{table:arguments sample}
\end{table}

\begin{table}[h]
    \centering
    \begin{tabular}{llp{8cm}}
        \toprule
        \textbf{Command} & \textbf{Default value} & \textbf{Description} \\
        \midrule
        \texttt{-d, --data} & \centering N/A & Filename of the input MSA. \\
        \texttt{-p, --path\_params} & \centering N/A & Path to the file containing the parameters of the DCA model. \\
        \texttt{-o, --output} & \centering N/A & Path to the folder where to save the output. \\
        \texttt{--alphabet} & \centering protein & Type of encoding for the sequences. Choose among \texttt{protein}, \texttt{rna}, \texttt{dna} or a user-defined string of tokens. \\
        \texttt{--device}\footnotemark[1] & \centering cuda & Device to be used between cuda (GPU) and CPU. Used in the Python version. \\
        \texttt{--dtype}\footnotemark[1] & \centering float32 & Data type to be used between float32 and float64. Used in the Python version. \\
        \bottomrule
    \end{tabular}
    \caption{Command line arguments for computing the DCA energies of an MSA.}
    \label{table:arguments energy}
\end{table}

\begin{table}[h]
    \centering
    \begin{tabular}{llp{8cm}}
        \toprule
        \textbf{Command} & \textbf{Default value} & \textbf{Description} \\
        \midrule
        \texttt{-d, --data} & \centering N/A & Filename of the input MSA containing the wild type. If multiple sequences are present, the first one is used. \\
        \texttt{-p, --path\_params} & \centering N/A & Path to the file containing the parameters of the DCA model. \\
        \texttt{-o, --output} & \centering N/A & Path to the folder where to save the output. \\
        \texttt{--alphabet} & \centering protein & Type of encoding for the sequences. Choose among \texttt{protein}, \texttt{rna}, \texttt{dna} or a user-defined string of tokens. \\
        \texttt{--device}\footnotemark[1] & \centering cuda & Device to be used between cuda (GPU) and CPU. Used in the Python version. \\
        \texttt{--dtype}\footnotemark[1] & \centering float32 & Data type to be used between float32 and float64. Used in the Python version. \\
        \bottomrule
    \end{tabular}
    \caption{Command line arguments for generating the deep mutational scan of a provided wild type sequence.}
    \label{table:arguments DMS}
\end{table}

\begin{table}[h]
    \centering
    \begin{tabular}{llp{8cm}}
        \toprule
        \textbf{Command} & \textbf{Default value} & \textbf{Description} \\
        \midrule
        \texttt{-p, --path\_params} & \centering N/A & Path to the file containing the parameters of the DCA model. \\
        \texttt{-o, --output} & \centering N/A & Path to the folder where to save the output. \\
        \texttt{-l, --label} & \centering None & If provided, adds a label to the output files inside the output folder.\\
        \texttt{--alphabet} & \centering protein & Type of encoding for the sequences. Choose among \texttt{protein}, \texttt{rna}, \texttt{dna} or a user-defined string of tokens. \\
        \texttt{--device}\footnotemark[1] & \centering cuda & Device to be used between cuda (GPU) and CPU. Used in the Python version. \\
        \texttt{--dtype}\footnotemark[1] & \centering float32 & Data type to be used between float32 and float64. Used in the Python version. \\
        \bottomrule
    \end{tabular}
    \caption{Command line arguments for computing the Frobenius contact matrix of a DCA model.}
    \label{table:arguments contacts}
\end{table}

\footnotetext[1]{These arguments are specific to the code implementation.}

\clearpage
\normalem

\bibliographystyle{unsrt}  
\bibliography{bibliography}  

\end{document}